\documentclass[norunningheads,11pt]{svmult}

\usepackage[sort&compress]{natbib}
\usepackage{bibentry}
\usepackage{fancyhdr}

\usepackage{times}
\usepackage{amsmath}
\usepackage{amssymb}
\usepackage{graphicx}
\usepackage{dcolumn}
\usepackage{bm}
\usepackage{makeidx}
\usepackage{multicol}

\usepackage[english]{babel}
\usepackage[latin2]{inputenc}
\usepackage[T1]{fontenc}

\def\ccc#1;#2{\left\langle #1 \left\vert #2 \right.\right\rangle}
\def\ev #1{\left\langle #1 \right\rangle}

\begin{document}
\pagestyle{empty}
\title*{Non-trivial scaling of fluctuations in the trading activity of the NYSE}
\author{J\'anos Kert\'esz \inst{1,2} \and Zolt\'an Eisler\inst{1}}
\institute{Department of Theoretical Physics, Budapest University of Technology and Economics, Budafoki \'ut 8, H-1111 Budapest, Hungary \and Laboratory of Computational Engineering, Helsinki University of Technology, P.O.Box 9203, FIN-02015 HUT, Finland}

{\Large\noindent\bf Non-trivial scaling of fluctuations in the trading}\vskip5.5pt
{\Large\noindent\bf activity of NYSE}\newline
\vskip7pt
\noindent J\'anos Kert\'esz$^{1,2}$ and Zolt\'an Eisler$^1$
\vskip7pt
\noindent {\small $^1$ Department of Theoretical Physics, Budapest University of Technology and Economics, Budafoki \'ut 8, H-1111 Budapest, Hungary}\newline
\noindent {\small $^2$ Laboratory of Computational Engineering, Helsinki University of Technology, P.O.Box 9203, FIN-02015 HUT, Finland}

\begin{abstract}
Complex systems comprise a large number of interacting elements, whose dynamics is not always a priori known. In these cases -- in order to uncover their key features -- we have to turn to empirical methods, one of which was recently introduced by Menezes and Barab\'asi. It is based on the observation that for the activity $f_i(t)$ of the constituents there is a power law relationship between the standard deviation and the mean value: $\sigma_i \propto \ev{f_i}^\alpha$. For stock market trading activity (traded value), good scaling over $5$ orders of magnitude with the exponent $\alpha = 0.72$ was observed. The origin of this non-trivial scaling can be traced back to a proportionality between the rate of trades $\ev{N}$ and their mean sizes $\ev{V}$. One finds $\ev{V} \propto \ev{N}^{0.69}$ for the $\sim1000$ largest companies of New York Stock Exchange. Model independent calculations show that these two types of scaling can be mapped onto each other, with an agreement between the error bars. Finally, there is a continuous increase in $\alpha$ if we look at fluctuations on an increasing time scale up to $20$ days.
\end{abstract}

\keywords{econophysics; stock market; fluctuation phenomena}\vskip20pt

\section{Introduction}

Although there is no generally recognized definition of complex systems, one of their widely accepted properties is that they comprise a large number of interacting constituents (or nodes) whose collective behavior forms spatial and/or temporal structures. Some of them are labeled "physical" because they are treated in the regular framework of physics. Nevertheless, the above scheme itself applies to a much wider range of systems, including the world economy consisting of companies that trade and compete. Human agents can interact with each other, e.g., by social networks or on the trading floor. We have little or no a priori knowledge about the laws governing these systems. Thus, very often our approach must be empirical. Recently, an increasing number of such systems have become possible to monitor through multichannel measurements. These offer the possibility to record and characterize the simultaneous time dependent behavior of many of the constituents. On the ground of these new datasets, an emerging technique (de Menezes and Barab\'asi 2004a) seems to be able to grasp important features of the internal dynamics in a model independent framework.

\section{Scaling of fluctuations in complex systems}

The method is based on a scaling relation that is observed for a growing range of systems:
The standard deviation $\sigma_i $ and time average $\ev{f_i}$ of the signal
$f_i(t)$ capturing the time dependent activity of elements $i=1,\dots,N$ follows the power law
\begin{equation}
\sigma_i \propto \ev{f_i}^\alpha ,
\label{eq:power-law}
\end{equation}
where we define 
\begin{equation}
\sigma_i = \sqrt{\ev{\left ( f_i - \ev{f_i}\right )^2}},
\end{equation}
and $\ev{\cdot}$ denotes time averaging.

This relationship is not unmotivated from a physicist's point of view. The constant $\alpha$ -- while not a universal exponent in the traditional sense -- is indeed the fingerprint of the microscopic dynamics of the system. Applications range from Internet traffic through river networks to econophysics. The latest advances (Menezes and Barab\'asi 2004b, Eisler and Kert\'esz 2005) have shown several possible scenarios leading to various scaling exponents:

\begin{enumerate}

\item The value $\alpha = 1$ always prevails in the presence of a \emph{dominant external driving force}. An example is web page visitation statistics. Here the main contribution to fluctuations comes from the fluctuating number of users surfing the web: a factor that is not intrinsic in the structure of the network. The situation is very similar for networks of roads or rivers.

\item There are systems, where the different mean activity of constituents comes exclusively from a different mean number of events. Individual events have the same mean contribution (impact) to a node's activity, only for more active nodes more of these events occur. When the \emph{central limit theorem} is applicable to the events, $\alpha = 1/2$. This behavior was observed for the logical elements of a computer chip and the data traffic of Internet routers.

\item Two mechanisms have been documented so far that can give rise to an intermediate value $1/2 < \alpha < 1$:
\begin{enumerate}
\item Because of the competition of external driving and internal fluctuations, it is possible that $\sigma$'s measured for finite systems display a crossover between $\alpha = 1/2$ and $\alpha = 1$ at a certain node strength $\ev{f}$. Then there exists an effective, intermediate value of $\alpha$, but actual scaling breaks down.
\item The other possibility is related to a very distinct character of internal dynamics: when elements with higher activity do not only experience more events, but those are also of larger impact. We call this property \emph{impact inhomogeneity}. Stock market trading belongs to this third group with $\alpha \approx 0.72$ for short time scales (see also Eisler et al. 2005).
\end{enumerate}
\end{enumerate}
In a recent model (Eisler and Kert\'esz 2005), the effect of impact inhomogeneity has been studied. Tokens are deposited on a Barab\'asi-Albert network (Albert and Barab\'asi 2002) and they are allowed to jump from node to node in every time step. Activity is generated when they arrive to a site. Every token that steps to a node $i$ generates an impact $V_i$ whose mean depends on the node degree $k_i$: $\ev{V_i}\propto k_i^\beta$. This gives rise to a scaling relation:
\begin{equation}
\ev{V_i}\propto\ev{N_i}^\beta.
\label{eq:inhom}
\end{equation}
The result of Eisler and Kert\'esz (2005) can then then be generalized as
\begin{equation}
\alpha = \frac{1}{2} \left (1+\frac{\beta}{\beta+1} \right ).
\label{eq:assume}
\end{equation}
Simulation results shown in Fig. \ref{fig:basic}(a) are in perfect agreement with formula \eqref{eq:assume}. This is an example that the value of $\alpha$ is basically determined by this impact inhomogeneity. If $\beta = 0$, i.e., the mean impact generated on all nodes is equal regardless of their degree, one recovers $\alpha = 1/2$. When $\beta > 0$, the events on more frequently visited nodes are also larger on average. Correspondingly, $\alpha > 1/2$.

\section{Application to stock market data}

Let us now turn to the case of the stock market. Data for the period $2000$--$2002$ was taken from the TAQ Database (New York Stock Exchange 2003). We define the activity $f_i(t)$ of stock $i$ as the capital flow in time windows of size $\Delta t$. In window $t$, $f_i(t)$ is the sum of $N_i(t)$ trading events. If we denote the value exchanged in the $n$'th trade of time window $t$ by $V_i(t;n)$, then the total traded value of stock $i$ is
\begin{equation}
f_i(t) = \sum_{n=1}^{N_i(t)}V_i(t;n).
\end{equation}
Then, $\ev{V}$ is the \emph{mean value per trade}, while $\ev{N}$ is the \emph{mean rate of trades}.

As we wish to calculate the mean and the standard deviation of this activity, it is essential that these quantities at least exist. Traded volumes and consequently traded values $f_i(t)$ are often considered to have a power law tail ($Prob(f>x) \propto x^{-\lambda}$) with an exponent $\lambda_i \sim 1.5-1.7$ (Gopikrishnan et al. 2000). This would imply, that the standard deviation is already divergent. Recent measurements, however, indicate that both of these quantities exist and that there is no unique $\lambda_i$ for a stock (Eisler and Kert\'esz unpublished).

\begin{figure}[!t]
\centerline{\includegraphics[height=131pt]{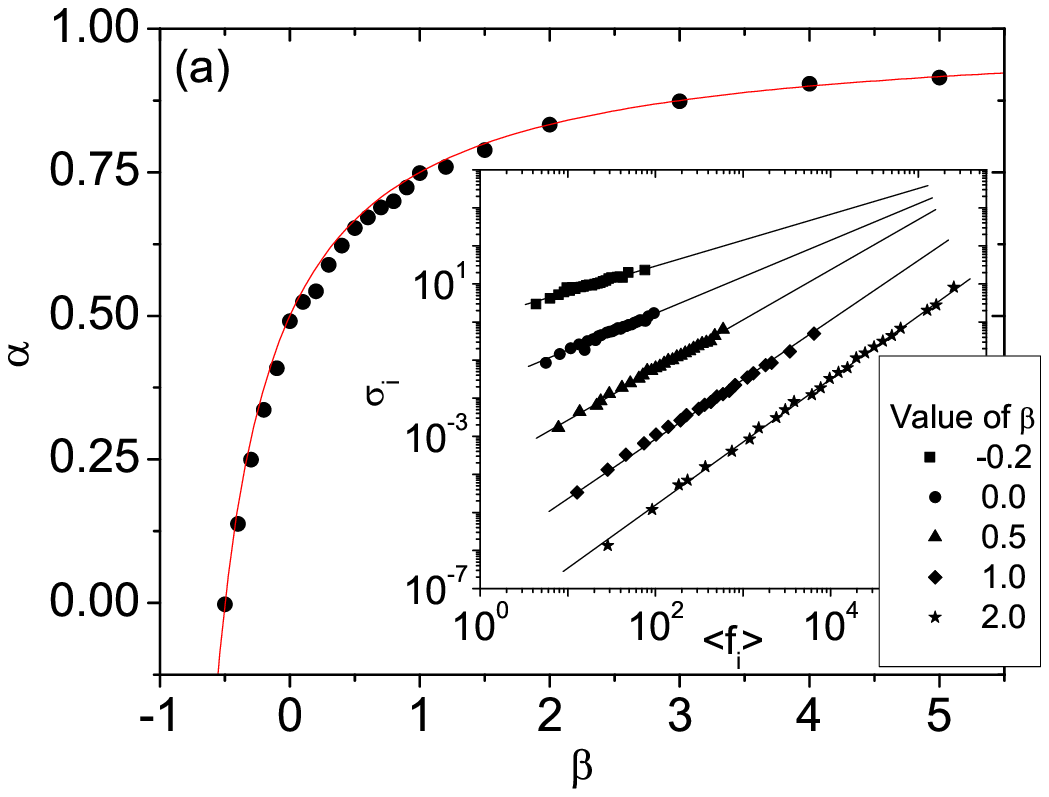}
\hskip10pt\includegraphics[height=130pt]{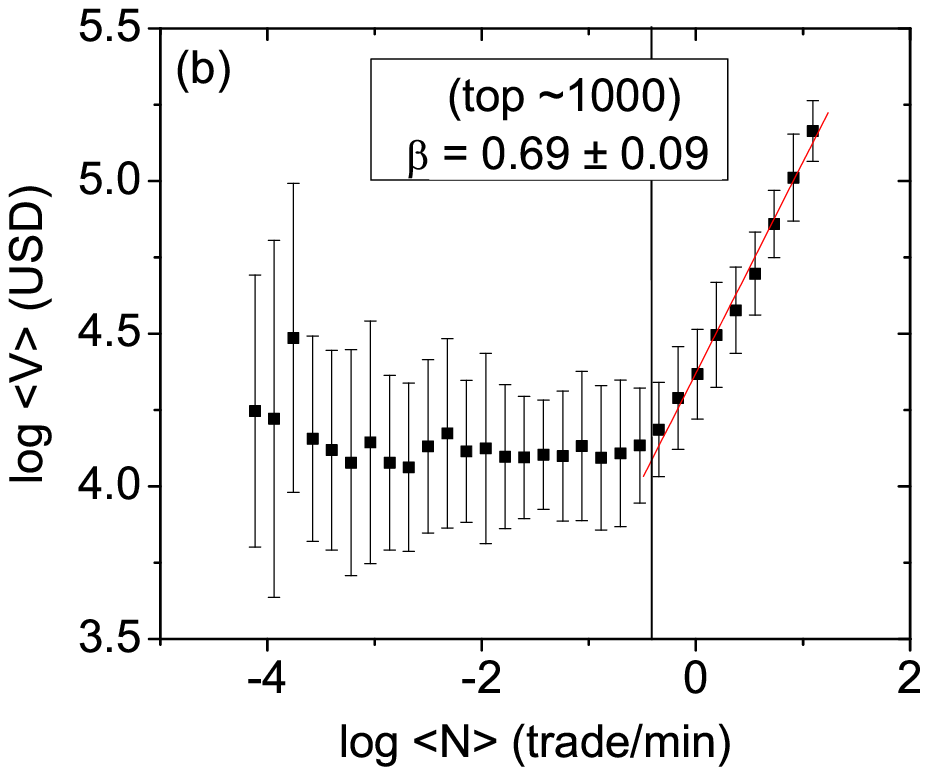}}
\caption{{\bf (a)} The value of $\alpha$ as a function of $\beta$ for the random walk model introduced by Eisler and Kert\'esz (2005). Circles give simulation results, while the solid line corresponds to \eqref{eq:assume}. The inset shows actual scaling plots for various values of $\beta$. {\bf (b)} Plot of mean value per trade $\ev{V}$ versus mean rate of trades $\ev{N}$ for NYSE. For smaller stocks there is no clear tendency. For the top $\sim1000$ companies, however, there is scaling with an exponent $\beta = 0.69 \pm 0.09$.}
\label{fig:basic}
\end{figure}

\begin{figure}[!t]
\centerline{\includegraphics[height=130pt]{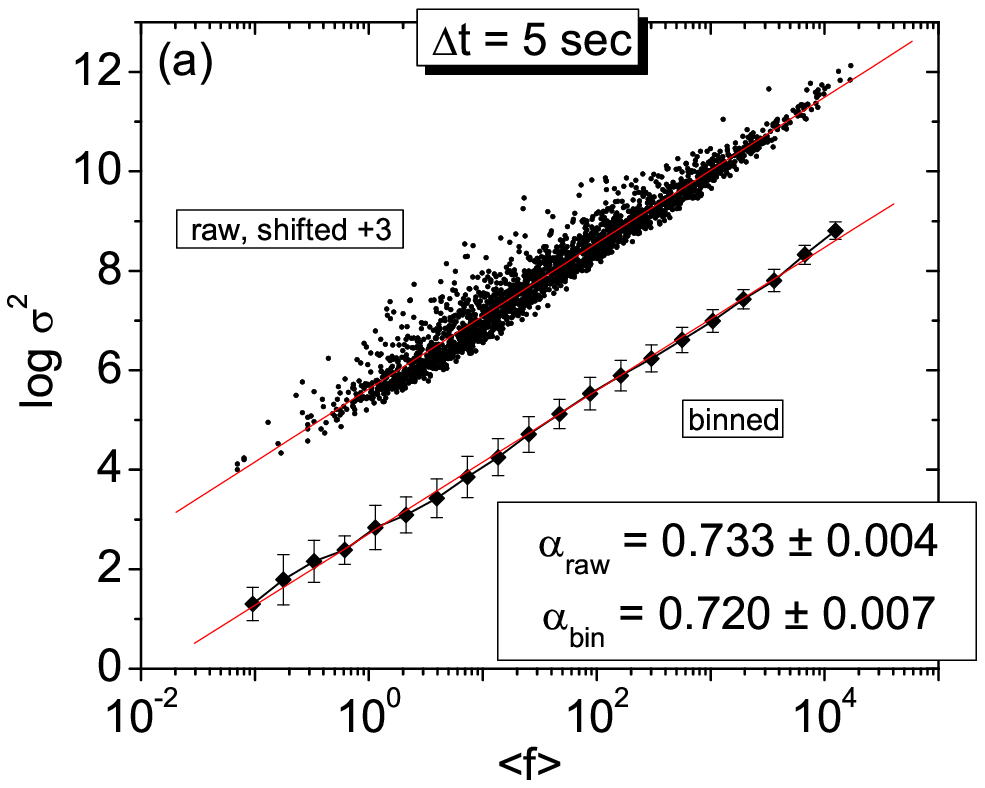}
\hskip4pt\includegraphics[height=132pt]{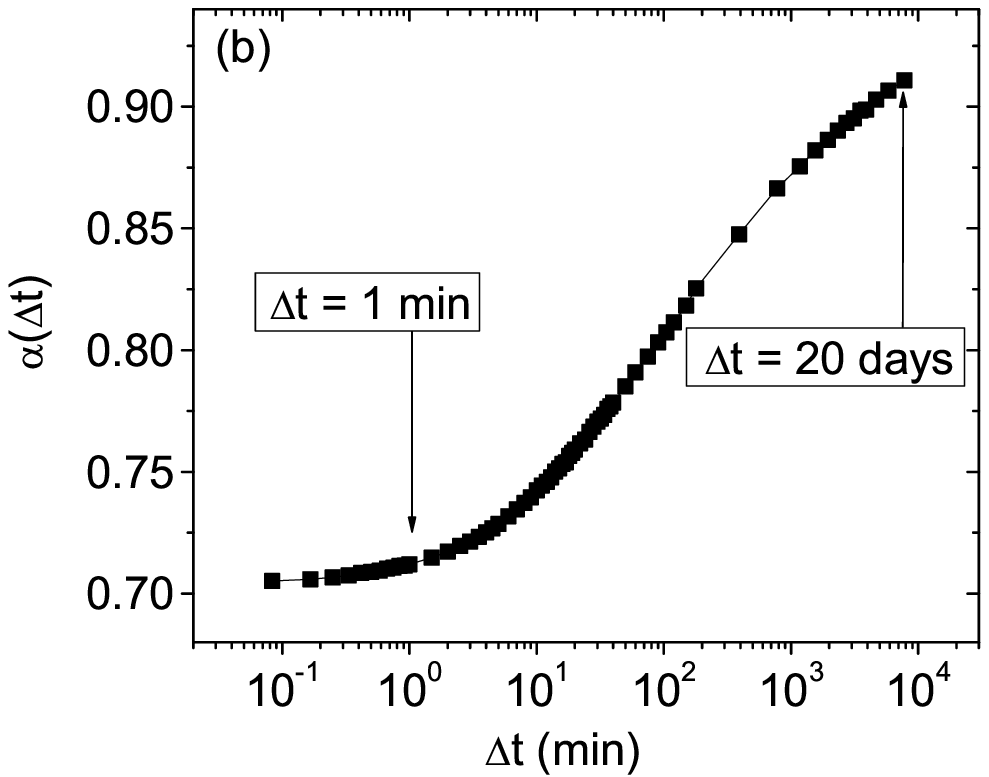}}
\caption{{\bf (a)} The scaling $\sigma \propto \ev{f}^\alpha$ for fluctuations of traded value at NYSE, $\Delta t = 5$ sec. Dots show raw results for each stock (shifted vertically for better visibility), the fitted slope is $\alpha_{raw}=0.733 \pm 0.004$. Diamonds show average $\sigma$'s for multiple stocks in a range of $\log \ev{f}$. This method corrects for bias that comes from the large number of stocks with low $\ev{f}$, one finds
$\alpha_{binned}=0.720 \pm 0.007$. {\bf (b)} The dependence of $\alpha$ on the time window $\Delta t$ for the NYSE data. One finds that up to $\Delta t = 1$ min, $\alpha \approx 0.72$, as expected from independent approximations. Then by increasing $\Delta t$, the value of $\alpha$ increases. This is due to the presence of strong autocorrelations in the activities $f(t)$ stemming from the clustering of trades.}
\label{fig:advanced}
\end{figure}

Then, it is possible to test the scaling relation \eqref{eq:power-law} and one finds good scaling over more than $5$ orders of magnitude in $\ev{f}$ with $\alpha \approx 0.72$. This is a value which can be -- at least partly -- explained in terms of impact inhomogeneity. We found\footnote{The result is qualitatively similar to those of Zumbach (2004) for the FTSE 100. He shows that both $\ev{N}$ and $\ev{V}$ scale as power laws with company capitalization for large companies. Capitalization dependence can be eliminated to recover \eqref{eq:inhom}.} that for the stocks of the $\sim1000$ largest companies of NYSE, $\beta = 0.69 \pm 0.09$ (see Fig. \ref{fig:basic}(b)). Substituting this into \eqref{eq:assume} we expect $\alpha = 0.71 \pm 0.01$, which is very close to the actual result seen from Fig. \ref{fig:advanced}(a). Note that although large error bars prevent us from testing \eqref{eq:assume} for smaller stocks, we still find that the scaling law \eqref{eq:power-law} holds. The exponent is unchanged, but this can only be explained by a detailed analysis of fluctuations.

The mechanism leading to non-trivial $\alpha$ via the scaling \eqref{eq:inhom} can be considered dominant only if the events are not strongly correlated. This condition is satisfied for short time windows $\Delta t$, when $\ev{N}\ll 1$. Interestingly, the value of $\alpha$ does not change noticably up to $\Delta t\sim1$ min. There is, however, another effect that is relevant to the value of $\alpha$ for longer time windows. For the NYSE data, $\alpha (\Delta t)$ increases continuously with $\Delta t$ (see Fig. \ref{fig:advanced}(b)). Previously (Eisler et al. 2005) this was attributed to the growing influence of external news: a kind of "driving". With longer time given for information to spread, the system was assumed to converge to the externally driven limit $\alpha = 1$. That mechanism would, however, lead to a \emph{crossover} to $\alpha = 1$ with increasing $\Delta t$ (Menezes and Barab\'asi 2004b). What is observed, is in fact \emph{not} a crossover. There is no breakdown of scaling as a function of $\ev{f}$ for intermediate $\Delta t$'s as one would expect between the regime of the two limiting exponents (Menezes and Barab\'asi 2004b). On the other hand, it is well known (see, e.g., Gopikrishnan et al. 2000), that the number of trades $N_i(t)$ is correlated. Individual trades tend to cluster together and this causes enhanced fluctuations in $N_i(t)$. This mechanism sets in at time windows for which the probability for two trades to coincide is no longer negligible. The scaling law \eqref{eq:power-law} itself is preserved, but the exponent $\alpha$ is strongly affected.

\section{Conclusions}

In the above we have outlined a recent type of scaling analysis for the fluctuations of activity in complex systems. We have shown that systems can be classified according to the scaling exponents $\alpha$. Then we have discussed how impact inhomogeneity and long range correlations give rise to non-trivial scaling exponents. Further research should clarify the interplay between fluctuations in the number of trades and in traded volumes/values in order to deepen the understanding of the market mechanism.

Acknowledgments: JK is member of the Center for Applied Mathematics and Computational Physics, BUTE. This research was supported by OTKA T049238. Thanks are due to A.-L. Barab\'asi and M.A. de Menezes.\hfill\vskip18pt

\noindent{\bf \large References}\vskip14pt
\noindent Albert R, Barab\'asi A-L (2002) Statistical mechanics of complex networks, Rev Mod
\indent Phys 74:47--97\newline
\noindent Eisler Z, Kert\'esz J (2005) Random walks on complex networks with inhomogeneous
\indent impact. arXiv:cond-mat/0501391, submitted to Phys Rev E\newline
\noindent Eisler Z, Kert\'esz J, Yook S-H, Barab\'asi A-L (2005) Multiscaling and non-universality
\indent in fluctuations of driven complex systems. Europhys Lett 69:664--670\newline
\noindent Gopikrishnan P, Plerou V, Gabaix X, Stanley HE (2000) Statistical properties of 
\indent share volume traded in financial markets. Phys Rev E 62:R4493-4496\newline
\noindent de Menezes MA, Barab\'asi A-L (2004a) Fluctuations in Network Dynamics. Phys
\indent Rev Lett 92:28701\newline
\noindent de Menezes MA, Barab\'asi A-L (2004b) Separating internal and external dynamics
\indent of complex systems. Phys Rev Lett 93:68701\newline
\noindent New York Stock Exchange (2003) The Trades and Quotes Database for 2000-2002.
\indent New York\newline
\noindent Zumbach G (2004) How the trading activity scales with the company sizes in the
\indent FTSE 100. arXiv:cond-mat/0407769, to appear in Quant Fin\newline


\end{document}